\documentclass[11pt]{article}
\usepackage{graphicx}
\begin{document}
\twocolumn[ \section*{\footnotesize{\it   Spacetime \& Substance}
\ {\bf 1}, \ {\rm no. \ 4, \ 172--175 \ (2000)\  \  \ \ (Invited
talk given at the Ukrainian Russian \\ \ \ \ conference \it
Gravitation, Cosmology and Relativistic Astrophysics,} \ {\rm 8-11
Nov., 2000, Kharkiv, Ukraine) \ \ \ \ \ \ \ \ \ \ \ \ \ \ \ \ \ \
\ \ \ \ \ \ \ \ \ \ \ \ \ \ \ \ \ \ \ \ \ }}

\bigskip\medskip

\section*{\bf  \ \ \ \ \ \ \ \ \ \ \
SPACE STRUCTURE AND QUANTUM MECHANICS}

\begin{center}
{\bf Volodymyr Krasnoholovets}
\end{center}

\begin{center}
{Institute of Physics, National Academy of Sciences, \\ Prospect
Nauky 46,   UA-03028 Ky\"{\i}v, Ukraine \ \
http://inerton.cjb.net}
\end{center}

{ \small A new concept of the constitution of nature is
considered. The constructed submicroscopic quantum mechanics is
deterministic and is characterised by elementary excitations of
the space net that is treated as the tesselation of balls, or
superparticles. Said excitations called "inertons" accompany any
canonical particle when it moves. It is shown theoretically that
the introduction of inertons obviates all conceptual difficulties
of orthodox quantum mechanics. The theory has been verified
experimentally. It is argued that just inertons play the role of
real carriers in the gravitational interaction.
\\

{\bf  Key words:} \ \  space, gravitation, inertons, quantum
mechanics
\\

{\bf PACS:} \ \ 03.65.Bz Foundations, theory of measurement,
miscellaneous theories. \\  03.75.-b Matter waves. \ \ 04.60.-m
Quantum gravity
\\
\medskip \vspace{1mm} } ]

\section{Introduction}

 \hspace*{\parindent}

 Sometimes this or that science one may treat
from not typical standpoints. For instance, in the mid-20th
century Arnold Toynbee [1] investigated historical studies
considering people history as a circulation of local
civilizations, which replace one another. 120 years ago Polish
linguist Michal Krasuski [2] was interested in the origin of
numerals: one, two, three, ..., ten, hundred, thousand, etc. The
study came him to the unexpected deduction: names of numerals were
associated with the every day fingers activity and names of
fingers in Ukrainian language. Thus he concluded that names of
numerals could be understood only from Ukrainian which, therefore,
is the most ancient language among all Aryan ones including
Sanscrit, Greek, Latin, etc.

Such kinds of studies are rather cognitive. But may quantum
physics be explored in a similar way?  Probably it may since
quantum physics is based on some initial notions (mass, particle,
quantization, particle energy $E=h\nu$, de Broglie wavelength
$\lambda=h/p$ and the matter waves, wave $\psi$-function and
long-range action, Compton wavelength $\lambda_{\rm Com}=h/mc$,
spin, fundamental constants, and so on), which all together have
never been treated in detail so far. A viewpoint of this type has
something in common with Louis de Broglie's, who used to say that
it is useful to reconsider the foundations of physics from time to
time. It is obvious that conducting such an analysis one will
touch both the foundations of quantum mechanics and the
foundations of quantum gravity since these two to be descent from
the same submicroscopic scale [3].

\section{The theory and results}
 \hspace*{\parindent}
We would start from the Dirac's remark [4] that the objections to
an aether posed by relativity were removed by quantum mechanics.
This means that a vague vacuum, or an empty space of general
relativity should make way for a substrate. The substrate cannot
directly be associated with an uncertain Higgs condensate of
models of grand unification of interactions (the condensate is not
constructed in a real 3D space). None the less the models of grand
unification basing on experimental results allow the calculation
of evolutions of three constants $\alpha_{\rm el.-magn.}$,
$\alpha_{\rm weak}$, and $\alpha_{\rm strong}$ as functions of
distance $r$. All the constants come together at $r\approx
10^{-28}$ cm (Fig. 1).

 At the same time  modern concepts of gravitational
interaction do not permit any similar analysis in principle. But
why? It is apparent that the main reason of such a distinguish of
the behaviour of gravitational interaction from the other
fundamental ones is caused by its initial phenomenological basis
while the detailed behaviour of the three other interactions were
constructed many years later resting on already well-developed
quantum mechanics. Those three interactions, electromagnetic,
weak, and strong are characterised by their own carriers, namely
photons, W$^{\pm}$ and Z bosons, and gluons, respectively. Hence
we may conclude that if someone tries to construct the
gravitational interaction starting from quantum mechanics, he may
also come to certain carriers, which will effect the direct
interaction between massive objects.
\begin{figure}
\begin{center}
\includegraphics[height=9.3cm, width=13.5cm]{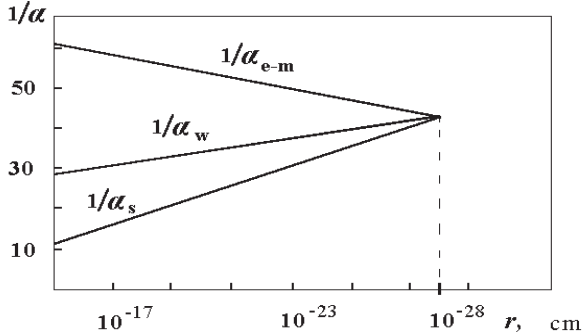}
\caption{Evolutional lines describing changes of the interaction
constants with distance.} \label{Figure 1}
\end{center}
\end{figure}
High energy physics has greatly advanced our knowledge about the
microstructure of real space and the structure of particles.
Specifically, unified models have proposed an abstract
"superparticle" whose different states are quarks, electron, muon,
neutrino, and others. Taking into account the facts mentioned
above a researcher whose specialty is condensed matter physics may
suggest that just those superparticles form a world substrate,
which shares discrete and continual properties. The introduction
of such a substrate in fundamental physics automatically implies
that both the relativity and the concept of unification of
interactions require a radical revision: 1) the gravitational
interaction will immediately be endowed by carriers, i.e., special
elementary excitations of the substrate; 2) the three other
interactions may not be elementary, particularly the weak and
strong ones -- they might result from the renormalization of some
kinds of more fundamental excitations of the substrate (such as
photons and inertons introduced below).

Let us construct a real space packing of superparticles leaning on
concepts and ideas used in condensed media physics. Such a
construction is in agreement with requirements of a mathematical
space [5]. The conceivable size of a superparticle may be equal to
$10^{-28}$ cm. Let superparticles being elastic densely put to
each other forming an entire substrate (quantum aether). The
substrate may be thought of as the degenerate space net. In the
theory proposed in papers [6-8] a local deformation of the space
net, i.e., a stable change of the initial volume of a
superparticle in the degenerate space net is associated with the
creation of a particle in it (Fig. 2).
\begin{figure}
\begin{center}
\includegraphics[scale=2.4]{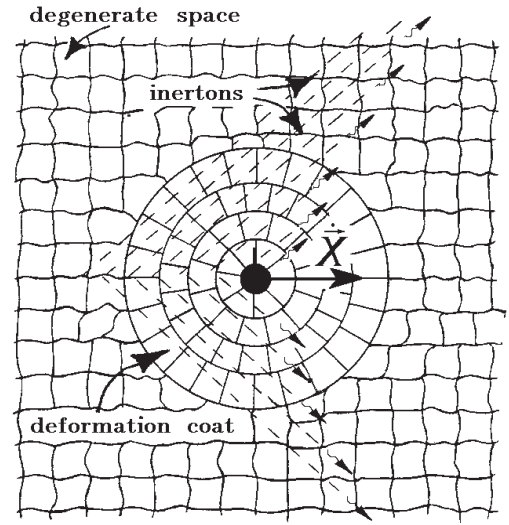}
\caption{The moving particle in the degenerate space net.}
\label{Figure 2}
\end{center}
\end{figure}
Unstable deformations constitute spatial excitations, or
quasi-particles, called "inertons" [6]. Any particle created in
the space net features its deformation coat (or singularity
region, or aether/space crystallite) which plays the role of a
screen shielding the particle from the degenerate space. The size
of the crystallite is equal to the Compton wavelength
$\lambda_{\rm Com}=h/Mc$ of the particle. Superparticles are
characterised by mass within this region and therefore possess
some properties typical for a solid crystal, for instance such as
collective vibrations. When the particle begins to move it
experiences friction striking superparticles. As a result of the
interaction with coming superparticles, the particle should emit
and then absorb elementary excitations, i.e. inertons [6-8]. Once
the particle moves running between fluctuating superparticles,
inertons migrate as typical quasi-particles, i.e., they carry bits
of space deformation hopping from superparticle to superparticle
by relay mechanism. The moving particle pulls its deformation
coat, or crystallite, since superparticles surrounding the
particle along all its path have time to be adjusted to the
particle organizing the coat.

The Lagrangian that characterises the motion of a particle is the
following [6]

\begin{eqnarray}\label{1}
L&=&{1\over {2}} g_{ij}{\dot X}^i(t){\dot X}^j(t) + {1\over
{2}}\sum^{N-1}_{l=0} {\tilde g}^{(l)}_{ij} {\dot
x}^i_{(l)}(t_{(l)}) {\dot x}^j_{(l)}(t_{(l)})    \nonumber   \\
&-&\sum_l^N [X^i {\dot x}^j_{(l)} + v_0\delta_{ij} x^j_{(l)}]
\delta_{t-\Delta t_{(l)}, t_{(l)}} \frac{\pi}{T_l}
\end{eqnarray}
Here, the first term describes the particle, the second term
describes the ensemble of inertons and the third one depicts the
interaction between the particle and inertons. $v_0$ is the
initial velocity of the particle, $\pi/T_l$ is the frequency of
collisions of the particle with inertons.

In the relativistic case we start from the Lagrangian
\begin{equation}\label{2}
  L_{\rm rel.} = - M_0 {\kern 1pt}c^2 \sqrt{1-v_0^2/c^2}
\end{equation}
in which the following transformation is made [7]
\begin{equation}\label{3}
  v_0^2 {\longrightarrow  {[g_{ij}\dot X^i \dot X^j +U(X,x,\dot
x)]/g}}.
\end{equation}
where the function $U$ is similar to the second and third terms in
expression (1).

The solution of the equations of motion has shown that the
particle oscillates along the trajectory: the particle velocity
changes periodically from $v_0$ to 0 and then from 0 to $v_0$ in
the interval of de Broglie wavelength $\lambda$. So from the
submicroscopic viewpoint the value $\lambda=h/Mv_0$ may be
remolded as the spatial period of the oscillatory moving particle.
The time period of the particle oscillations is $T=\lambda/v_0$. A
similar relation is true for the inerton cloud enclosing the
particle. The cloud oscillations are specified by amplitude
$\Lambda=cT$ where $c$ is the velocity of inertons in the
degenerate space net, which might be equal to the velocity of
light. The amplitude $\Lambda$ is connected with the de Broglie
wavelength by the relationship [6]
\begin{equation}\label{4}
  \Lambda=\lambda c/v_0.
\end{equation}

It is proved that the motion of the particle is deterministic and
features the relationships
\begin{equation}\label{5}
E=h\nu, \ \ \ \ \ \ \ \lambda=h/Mv_0
\end{equation}
where $E=Mv_0^2/2$ and $M=M_0/\sqrt{1-v_0^2/c^2}$. As known [9]
just the availability of relationships (5) allows the introduction
of the Schr\"odinger formalism, which due to relationship (4) is
correct in the region covered by the distance $\Lambda$ from the
particle. Thus the introduction of inertons automatically removes
lang-range action from any quantum system restricting the effect
of the Schr\"odinger formalism by the value of amplitude
$\Lambda$.

The problem of spin is solved [8] by the introduction of one more
degree of freedom in expression (3), i.e., we add a new matrix
form function $U_{\alpha}$ (where $\alpha=\uparrow,  \downarrow$)
to the function $U$. The matrix describes two possible pulsations
of the particle: ahead ($\uparrow$) and back ($\downarrow$) in
relation to the vector of particle's motion velocity. This sheds
light [8] on the inner reason of transformation of the total
Hamiltonian of a particle
\begin{equation}\label{6}
H^{\rm part. tot}= \sqrt{c^2\vec p^{\ 2} +c^2\vec\pi^{\
2}_{\uparrow (\downarrow)} + M_0^2c^4}
\end{equation}
to the Dirac Hamiltonian operator
\begin{equation}\label{7}
\hat H_{\rm Dirac}=c{\kern 1pt}\hat{\vec{\alpha}}{\kern 1pt}
\hat{\vec p} +{\hat{\varrho}}_{\kern 1pt 3}M_0{\kern 1pt}c^2.
\end{equation}

The theory developed yields the very interesting relationship [8],
namely it groups together the amplitude of inerton cloud with the
Compton wavelength
\begin{equation}\label{8}
\Lambda = \tilde\lambda_{\rm Com}{\kern 1pt}c^2/v_0^2.
\end{equation}
With $v_0^2/c^2<<1$, the inerton cloud amplitude
$\Lambda>>\lambda_{\rm Com}$ and the inerton cloud governs the
motion of the particle, as already at a distance of $\Lambda$ from
the particle the cloud undergoes obstacles and passes the
corresponding information to the particle. This is the de Broglie
"motion by guidance" and the utilization of the Schr\"odinger
formalism is quite correct in this case.

In the approximation $v_0 \rightarrow c$, $\Lambda \approx
\lambda_{\rm Com}$ and therefore the cloud of inertons completely
closed in the crystallite surrounding the particle. This case
falls under the Dirac formalism.

Thus inertons surrounding a moving particle make up a substructure
of the matter waves, which so far have been treated only in the
framework of the wave $\psi$-function probabilistic formalism and
any physical interpretation has not been taken into account.

\section{Experimental corroboration}
 \hspace*{\parindent}
The theory of submicroscopic quantum mechanics has been verified.
Paper [10] demonstrates how inerton cloud expanded around moving
electrons manifest themselves in numerous experiments. The paper
deals with experimental and theoretical results available when
laser-induced gas ionisation phenomena and photoemission from a
laser-irradiated metal take place.

In work [11] the impact of inertons on the collective behaviour of
atoms in a solid has theoretically been treated and then
experimentally approved in metal specimens. It has been derived
that the force matrix $W_{\alpha \beta}$ that determines three
branches of acoustic vibrations in the crystal lattice consists of
two components
\begin{equation}\label{9}
  W_{\alpha\beta}(\vec k) = \tilde V_{\alpha \beta}(\vec k) + \tilde
\tau^{-1}_{\alpha \beta} \sum_{\alpha^\prime} \tilde
\tau^{-1}_{\alpha^\prime \beta} (\vec k) {e_{\alpha^\prime}\over
e_{\beta}}
\end{equation}
Here, $\tilde V_{\alpha \beta}$ is the usual term caused by the
elastic electromagnetic interaction of atoms and the second term
is originated from the overlapping of inerton clouds of adjacent
atoms. The availability of the second term means that an outside
inerton field is able to influence the crystal lattice increasing
amplitudes of vibrating atoms. The experiment which assumes the
presence of the hypothetical inerton field has been performed. The
terrestrial globe has been considered as a source of inertons. The
expected changes in the structure of test metal specimens caused
by the Earth inerton field in fact have been convincingly fixed in
electron micrographs [11].

Moreover just recently the theory has been tested for truth in the
experiment on the hydrogen atoms clustering in the
$\delta$-KIO$_3\cdot$HIO$_3$ crystal [12]. We have considered the
cluster formation of atoms in a model when the potentials of
attraction and repulsion are parted from one another. Proceeding
from submicroscopic quantum mechanics we come to the following
form of two parts of the Lennard-Jones potential
\begin{equation}\label{10}
V_{\rm att}(r)= -\epsilon {\kern 1pt}\big( \frac {g}{r} \big)^6 +
\frac 12 \gamma r^2; \ \ \ \ V_{\rm rep}(r)= \epsilon {\kern
1pt}\big( \frac {g}{r} \big)^{12}.
\end{equation}
Here, the small correction $\gamma r^2/2$ is stipulated by an
elastic response of the space net on the motion of acoustic
excitations, i.e. phonons, in the crystal lattice. The calculated
number of hydrogen atoms in a cluster equals
\begin{equation}\label{11}
{\cal N}\simeq \Big( { {3{\kern 1pt}\epsilon}\over {\gamma g^2}}
\Big)^{3/5}
\end{equation}
where $\epsilon$ and $g$ are the bound energy and the length of
O--H bond respectively and $\gamma$ is the inerton elasticity
constant of the cluster. The  IR spectra obtained yield the
reliable evidence of the cluster formation in the crystal studied
(${\cal N}=24$, the number of cooperated hydrogen atoms [12]).

\section{Conclusion}
 \hspace*{\parindent}
Summarizing we can infer that just the inerton field, {\bf a new
physical field}, whose carriers -- inertons --  make a
substructure of the matter waves, generates the quantum mechanics
formalism in the region from $10^{-28}$ cm to the atom size. The
radius of action of the field of a particle is limited by the
amplitude $\Lambda$ (4) of particle's inerton cloud. This also
signifies that the gravitational radius of a particle is
restricted by the same distance $\Lambda$ since any piece of
information about the particle cannot be found beyond the bounds
of its inerton cloud. Thus the dynamic inerton field becomes a
real candidate for the understanding the gravitation phenomenon
and yet the research conducted denies an option of the existence
of gravitons of general relativity without any doubt. The inerton
field is capable also to account for macroscopic phenomena
trespassing upon the range traditionally describing by general
relativity.  This means that general relativity loses its monopole
rights of the all-embracing theory: the static relativity should
be replaced for a dynamic theory based on the inerton field that
realizes the direct interaction between massive objects. However
this is the other problem which is still waiting for the solution.

\end{document}